\documentstyle[11pt]{article}

\makeatletter
\@addtoreset{equation}{section}
\makeatother

\let\a=\alpha

\let\la=\label  
   
 \def\bd{\begin{document}} \def\ed{\end{document}}
\def\ds{\documentstyle} \let\fr=\frac \let\bl=\bigl \let\br=\bigr
\let\Br=\Bigr \let\Bl=\Bigl 
\let\bm=\bibitem
\let\na=\nabla
\let\pa=\partial \let\ov=\overline 
\newcommand{\be}{\begin{equation}} 
\newcommand{\ee}{\end{equation}} 
\def\ba{\begin{array}}
\def\ea{\end{array}}
\def\ft#1#2{{\textstyle{{\scriptstyle #1}\over {\scriptstyle #2}}}}
\def\fft#1#2{{#1 \over #2}}
\def\del{\partial}
\def\vp{\varphi}
\def\sst#1{{\scriptscriptstyle #1}}
\def\oneone{\rlap 1\mkern4mu{\rm l}}
\def\td{\tilde}
\def\wtd{\widetilde}
\newcommand{\ho}[1]{$\, ^{#1}$}
\newcommand{\hoch}[1]{$\, ^{#1}$}
\newcommand{\bea}{\begin{eqnarray}} 
\newcommand{\eea}{\end{eqnarray}} 
\newcommand{\ra}{\rightarrow}
\newcommand{\lra}{\longrightarrow}
\newcommand{\Lra}{\Leftrightarrow}
\newcommand{\ap}{\alpha^\prime}
\newcommand{\bp}{\tilde \beta^\prime}
\newcommand{\tr}{{\rm tr} }
\newcommand{\Tr}{{\rm Tr} } 
\newcommand{\NP}{Nucl. Phys. }
\newcommand{\tamphys}{\it Center for Theoretical Physics\\
Texas A\&M University, College Station, Texas 77843}
\newcommand{\auth}{M. J. Duff\hoch{\dagger}, H.
L\"u\hoch{\ddagger} and C. N. Pope\hoch{\ddagger}}

\begin{document}

\hfill{CTP-TAMU-14/96}

\hfill{hep-th/9604052}

\vspace{20pt}

\begin{center}
{ \large {\bf The Black Branes of M-theory}}

\vspace{30pt}

\auth

\vspace{15pt}

{\tamphys}

\vspace{40pt}

\underline{ABSTRACT}
\end{center}

   We present a class of black $p$-brane solutions of M-theory which were
hitherto known only in the extremal supersymmetric limit, and calculate their
macroscopic entropy and temperature. 

{\vfill\leftline{}\vfill
\vskip	10pt
\footnoterule
{\footnotesize
	\hoch{\dagger}	Research supported in part by NSF Grant	PHY-9411543
\vskip	-12pt} \vskip	10pt {\footnotesize
	\hoch{\ddagger}	Research supported in part by DOE 
Grant DE-FG05-91-ER40633 \vskip	-12pt}}

\pagebreak
\setcounter{page}{1}

\section{Introduction}
\la{Introduction}

     There is now a consensus that the best candidate for a unified theory
underlying all physical phenomena is no longer ten-dimensional string theory
but rather eleven-dimensional {\it M-theory} .  The precise formulation of
M-theory is unclear but membranes and fivebranes enter in a crucial way,
owing to the presence of a $4$-form field strength $F_4$ in the
corresponding eleven-dimensional supergravity theory \cite{cjs}.  The
membrane is characterized by a tension $T_3$ and an ``electric'' charge
$Q_3=\int_{S^7}*F_4$.   For $T_3>Q_3$, the membrane is ``black''
\cite{guven}, exhibiting an outer event horizon at $r=r_+$ and an inner
horizon at $r=r_-$, where $r=\sqrt{y^my_m}$ and where $y^m$, $m=1,2,...,8$,
are the coordinates transverse to the membrane.   In the extremal
tension=charge limit, the two horizons coincide, and one recovers the
fundamental supermembrane solution which preserves half of the spacetime
supersymmetries \cite{ds}.  This supermembrane admits a covariant
Green-Schwarz action \cite{bst}.   Similar remarks apply to the fivebrane 
which is characterized by a tension $T_6$ and ``magnetic charge''
$P_6=\int_{S^4}F_4$.  It is also black when $T_6>P_6$ and also preserves
half the supersymmetries in the extremal limit \cite{guven}.  There is, to
date, no covariant fivebrane action, however. Upon compactification of
M-theory to a lower spacetime dimension, a bewildering array of other black
$p$-branes make their appearance in the theory, owing to the presence of a
variety of $(p+2)$-form field strengths in the lower-dimensional
supergravity theory \cite{hs,dl}.  Some of these $p$-branes may be
interpreted as reductions of the eleven-dimensional ones or wrappings of the
eleven-dimensional ones around cycles of the compactifying manifold
\cite{dkrm,ht,dkl,t2}.  In particular, one may obtain as special cases the
four-dimensional black holes ($p=1$). It has been suggested that, in the
extremal limit, these black holes may be identified with BPS saturated string
states \cite{dr,sen1,km,peet}. Moreover, it is sometimes the case that
multiply-charged black holes may be regarded as bound states at threshold of
singly charged black holes \cite{dr,sen1,dlr,r}.  Apart from their
importance in the understanding of M-theory, therefore, these  black
$p$-branes have recently come to the fore as a way of providing a
microscopic explanation of the Hawking entropy and temperature formulae
[17-28] which have long been something of an enigma . This latter progress
has been made possible by the recognition that some $p$-branes carrying
Ramond-Ramond charges also admit an interpretation as Dirichlet-branes, or
$D$-branes, and are therefore amenable to the calculational power of
conformal field theory \cite{p}. 

    The compactified eleven-dimensional supergravity theory admits a
consistent truncation to the following set of fields: the metric tensor
$g_{\sst{MN}}$, a set of $N$ scalar fields $\vec \phi=(\phi_1, \ldots,
\phi_{\sst N})$, and $N$ field strengths $F_\a$ of rank $n$.  The Lagrangian
for these fields takes the form  \cite{lpsol,lpmulti} 
\be 
e^{-1} {\cal L} = R -\ft12 (\del
\vec\phi)^2 -\fft1{2n!} \sum_{\a=1}^N e^{\vec a_\a\cdot \vec \phi} F_\a^2
\ ,\label{mslag}  
\ee 
where $\vec a_\a$ are constant vectors characteristic of the supergravity 
theory.  The purpose of the present paper is to display a universal class of
(non-rotating) black $p$-brane solutions to (\ref{mslag}) and to calculate their
classical entropy and temperature.
 
       As discussed in section 2, it is also possible to make a further
consistent truncation to a single scalar  $\phi$ and single field strength
$F$: 
\be 
e^{-1} {\cal L} = R - \ft12 (\del\phi)^2 -\fft1{2n!} e^{a\phi}
F^2 \ ,\label{sslagi} 
\ee
where the parameter $a$ can be conveniently re-expressed as
\be
a^2 = \Delta - \fft{2 d \td d}{D-2}\ ,
\ee
since $\Delta$ is a parameter that is preserved under dimensional
reduction \cite{lpss}.  Special solutions of this theory have been
considered before in the literature.  Purely electric or purely magnetic
black $p$-branes were considered in \cite{hs} for $D=10$ dimensions and in
\cite{dl} for general dimensions $D \leq 11$ .  All these had $\Delta=4$. In
the case of {\it extremal} black $p$-branes, these were generalized to other
values of $\Delta$ in \cite{lpss,lpsol}.  Certain non-extremal non-dilatonic 
($a=0$) black $p$-branes were also obtained in \cite{ght}

      A particularly interesting class of solutions are the {\it dyonic}
$p$-branes.  Dyonic $p$-brane occur in dimensions $D=2n$, where the
$n$-index  field strengths can carry both electric and magnetic charges. 
There are two types of dyonic solution. In the first type, each individual
field strength in (\ref{mslag}) carries either electric charge or magnetic 
charge, but not both. A particularly interesting example, owing to its
non-vanishing entropy even in the extremal limit \cite{lw}, is provided by
the four-dimensional dyonic black hole. This is the $a=0$
(Reissner-Nordstrom) solution, recently identified as a solution of
heterotic string theory \cite{dr}, but known for many years to be a solution
of M-theory \cite{dp,po}.   The construction of black dyonic $p$-branes of
this type is identical to that for the solutions with purely electric or
purely magnetic charges, discussed in section (3). 
 
      In section (4), we shall construct black dyonic $p$-branes of the
second type, where there is one field strength, which carries both electric
and magnetic charge.  Special cases of these have also been considered
before: the self-dual threebrane in $D=10$ \cite{hs,dl2}, the extremal
self-dual string \cite{dl} and extremal dyonic string in $D=6$ \cite{dfkr}, a
black self-dual string in $D=6$ \cite{ght,hs2} and a different dyonic black
hole in $D=4$ \cite{lpsol}. See also \cite{cy} for the most general
spherically symmetric extremal dyonic black hole solutions of the toroidally
compactified heterotic string. 

      Black multi-scalar $p$-branes, the extremal limits of which may be
found in \cite{lpmulti}, are discussed in section (5). 
     
       The usual form of the metric for an isotropic $p$-brane in $D$
dimensions is given by 
\be
ds^2 =e^{2A} (-dt^2 + dx^idx^i) + e^{2B} (dr^2 + r^2 d\Omega^2)\ ,
\label{isometric}
\ee
where the coordinates $(t, x^i)$ parameterise the $d$-dimensional
world-volume of the $p$-brane.  The remaining coordinates of the $D$
dimensional spacetime are $r$ and the coordinates on a $(D-d-1)$-dimensional
unit sphere, whose metric is $d\Omega^2$.  The functions $A$ and $B$ depend
on the coordinate $r$ only, as do the dilatonic scalar fields.  The field
strengths $F_\a$ can carry either electric or magnetic charge, and are given
by 
\be
F_\a = \lambda_\a\, *\!\epsilon_{\sst D-n}\ ,\qquad {\rm or}\qquad
F_\a = \lambda_\a \, \epsilon_n\ ,\label{fansatz}
\ee
where $\epsilon_n$ is the volume form on the unit sphere $d\Omega^2$.  The 
former case describes an elementary $p$-brane solution with $d=n-1$ and 
electric charge $\lambda_\a= Q_\a$; the latter a solitonic $p$-brane
solution with $d=D-n-1$ and magnetic charge $\lambda_\a=P_\a$. 

      Solutions of supergravity theories with metrics of this form include
extremal supersymmetric $p$-brane solitons, which saturate the Bogomol'nyi
bound.  The mass per unit $p$-volume of such a solution is equal to the sum
of the electric and/or magnetic charges carried by participating field
strengths.   More general classes of ``black'' solutions exist in which the
mass is an independent free parameter.  In this paper, we shall show that
there is a universal recipe for constructing such non-extremal
generalisations of $p$-brane solutions, in which the metric
(\ref{isometric}) is replaced by 
\be
ds^2 =e^{2A}(-e^{2f} dt^2 + dx^idx^i) + e^{2B}(e^{-2f}dr^2 +
r^2 d\Omega^2)\ .\label{blackmetric}
\ee
Like $A$ and $B$, $f$ is a function of $r$.  The ans\"atze for the field
strengths (\ref{fansatz}) remain the same as in the extremal case.
Remarkably, it turns out that the functions $A$, $B$ and $\vec \phi$ take
exactly the same form as they do in the extremal case, but for rescaled
values of the electric and magnetic charges.  The function $f$ has a
completely universal form: 
\be 
e^{2f} = 1 - \fft{k}{r^{\td d}}\ ,\label{fsol}
\ee
where $\td d = D-d-2$.  If $k$ is positive, the metric has an outer event
horizon at $r=r_+=k^{1/\td d}$.  When $k=0$, the solution becomes extremal,
and the horizon coincides with the location of the curvature singularity at
$r=0$. 

     The temperature of a black $p$-brane can be calculated by examining the
behaviour of the metric (\ref{blackmetric}) in the Euclidean regime in the
vicinity of the outer horizon $r=r_+$.  Setting $t=i\tau$ and $1-k r^{-\td
d}= \rho^2$, the metric (\ref{blackmetric}) becomes 
\be
ds^2=\fft{4r_+^2}{\td d^2} e^{2B(r_+)} \Big (d\rho^2 + 
\fft{\td d^2}{4 r_+^2} e^{2A(r_+) - 2B(r_+)} \rho^2 d\tau^2 + \cdots
\Big ) \ .
\ee
We see that the conical singularity at the outer horizon ($\rho=0$) is
avoided if $\tau$ is assigned the period $(4\pi r_+/\td d) e^{B(r_+) -
A(r_+)}$.  The inverse of this periodicity in imaginary time is the Hawking
temperature, 
\be 
T = \fft{\td d}{4\pi r_+} e^{A(r_+) - B(r_+)}\ .\label{temp}
\ee
We may also calculate the entropy per unit $p$-volume of the black
$p$-brane, which is given by one quarter of the area of the outer horizon.
Thus we have 
\be
S=\ft14 r_+^{\td d +1} e^{(\td d+1) B(r_+)+(d-1)A(r_+)} 
\omega_{\td d +1}\ ,\label{entropy}
\ee
where $\omega_{\td d+1} = 2\pi^{\td d/2 + 1}/(\ft12\td d)!$ is the volume of
the unit $(\td d+1)$-sphere 

     In subsequent sections, we shall generalise various kinds of extremal
$p$-brane solutions to obtain black single-scalar elementary and solitonic
$p$-branes, black dyonic $p$-branes and black multi-scalar $p$-branes. The
metric ansatz (\ref{blackmetric}) gives rise to non-isotropic $p$-brane
solutions for $d\ge 2$, in the sense that the Poincar\'e symmetry of the
$d$-dimensional world volume is broken.  When $d=1$, however, the black hole
solutions remain isotropic.  In the extremal black hole solutions, the
quantity $dA + \td d B$ vanishes, where $A$ and $B$ are defined in
(\ref{isometric}); whilst in the non-extremal cases, this quantity is
non-vanishing.  Isotropic $p$-brane solutions with $dA + \td d B\ne 0$ were
discussed in \cite{lpx}. 

\section{Single-scalar black $p$-branes}

      The Lagrangian (\ref{mslag}) can be consistently reduced to a
Lagrangian for a single scalar and a single field strength
\be
e^{-1} {\cal L} = R - \ft12 (\del\phi)^2 -\fft1{2n!} e^{a\phi} F^2
\ ,\label{sslag}
\ee
where $a$, $\phi$ and $F$ are given by \cite{lpsol}
\bea
a^2 &=& (\sum_{\a,\beta} (M^{-1})_{\a\beta})^{-1}\ ,\qquad
\phi=a\sum_{\a,\beta} (M^{-1})_{\a\beta}\, \vec a_\a\cdot\vec \phi\ ,
\nonumber\\
(F_\a)^2 &=& a^2 \sum_{\beta} (M^{-1})_{\a\beta}\, F^2\ ,\label{mtsred}
\eea
and $M_{\a\beta}= \vec a_\a \cdot \vec a_\beta$. The parameter $a$ can 
conveniently be re-expressed as 
\be
a^2 = \Delta - \fft{2 d \td d}{D-2}\ ,
\ee
where $\Delta$ is a parameter that is preserved under dimensional reduction
\cite{lpss}.  Supersymmetric $p$-brane solutions can arise only when the
value of $\Delta$ is given by $\Delta=4/N$, with $N$ field strengths
participating in the solution.  This occurs when the dot products of the
dilaton vectors $\vec a_\a$ satisfy \cite{lpmulti} 
\be
M_{\a\beta} = 4 \delta_{\a\beta} - \fft{2 d \td d}{D-2}\ .\label{mmatrix}
\ee
An interesting special case is provided by the four-dimensional black holes
with $a^2=3,1,1/3,0$, i.e $N=1,2,3,4$ whose extremal limits admit the
interpretation of $1,2,3,4$-particle bound states at threshold
\cite{dr,dlr,r}. Their $D=11$ interpretation has recently been discussed in
\cite{pt,t1}. 
     
       To begin, let us consider the more general metric
\be
ds^2 =-e^{2u} dt^2 + e^{2A} dx^idx^i + e^{2v} dr^2 + e^{2B} r^2 d\Omega^2\ .
\label{mgen}
\ee
It is straightforward to show that the Ricci tensor for this metric has the 
following non-vanishing components
\bea
R_{00} &=& e^{2(u - v)} \Big( u'' - u'v' + v'^2 + (d-1) u'
A' + (\td d+1) u'(B' + \fft1{r}) \Big)\ ,\nonumber\\ 
R_{ij} &=&- e^{2(A-v)} \Big ( A'' -A'v' + A' u' + (d-1) A'^2 + 
(\td d+1) A'(B' + \fft{1}{r})\Big ) \delta_{ij}\ ,\nonumber\\
R_{rr} &=& -u'' + u'v' - u'^2- (d-1) A'' + (d-1) A' v' - (d-1) A'^2 -
(\td d +1) B''\ ,\nonumber\\
&& + \fft{\td d+1}{r} v' - \fft{2(\td d +1)}{r} B' + (\td d+1)v' B' -
(\td d + 1) B'^2 \ ,\label{ricci}\\
R_{ab} &=& - e^{2(B-v)} \Big( B'' + (B' +\fft{1}{r}) \big[u' -v' + (d-1) A'
+ (\td d+ 1) (B' +\fft{1}{r})\big] - \fft{1}{r^2} \Big) g_{ab} + \td d
g_{ab} \ ,\nonumber 
\eea
where a prime denotes a derivative with respect to $r$, and $g_{ab}$ is the 
metric on the unit $(\td d+1)$-sphere.  For future reference, we note that 
the ADM mass per unit $p$-volume for this metric is given by \cite{l}
\be
m= \Big[ (d-1) (e^{2A})' r^{\td d+1} + (\td d +1) (e^{2B})'
r^{\td d+1} - (\td d+1) (e^{2v} - e^{2B}) r^{\td d}\,
\Big] \Big|_{r\rightarrow \infty}\ .\label{massfor}
\ee

     The Ricci tensor for the metric (\ref{blackmetric}) is given by
(\ref{ricci}) with $u=2(A+f)$ and $v=2(B-f)$.  As in the case of isotropic
$p$-brane solutions, the equations of motion simplify dramatically after
imposing the ansatz 
\be
dA + \td d B=0\ .\label{drama}
\ee  
Furthermore, the structure of the equations of motion implies that it is
natural to take 
\be
f'' + \fft{\td d +1}{r} f' + 2 f'^2 =0\ ,\label{feq}
\ee
which has the solution given by (\ref{fsol}).  Note that we have chosen the
asymptotic value of $f$ to be zero at $r=\infty$.  This is necessary in
order that the metric (\ref{blackmetric}) be Minkowskian at $r=\infty$. The
equations of motion then reduce to the following three simple equations: 
\bea 
\phi'' + \fft{\td d+1}{r} \phi' + 2 \phi' f' &=& -\fft{\epsilon a}{2} s^2
e^{-2f}\ , \nonumber\\ 
A'' + \fft{\td d+1}{r} A' + 2 A' f' &=& \fft{\td d}{2(D-2)} s^2 e^{-2f}\ ,
\label{eom1}\\
d(D-2) A' + \ft12{\td d} \phi'^2 + 2(D-2)A'f' &=& \ft12{\td d} s^2 e^{-2f}
\ ,\nonumber
\eea
where $s$ is given by
\be
s=\lambda e^{-\ft12\epsilon a \phi+ dA}\, r^{-(\td d+1)}\ ,\label{seq}
\ee
and $\epsilon=1$ for elementary solutions and $\epsilon=-1$ for solitonic 
solutions. The last equation in (\ref{eom1}) is a first integral of the
first two equations, and hence determines an integration constant.  The
first two equations in (\ref{eom1}) imply that we can naturally solve for
the dilaton $\phi$ by taking $\phi =  a (D-2) A/\td d$.  The remaining
equation can then be easily solved by making the ansatz that the function
$A$ takes the identical form as in the extremal case, but with a rescaled
charge, {\it i.e.}\ it satisfies 
\be
A'' + \fft{\td d+1}{r} A' = \fft{\td d}{2(D-2)} \wtd s^2\ ,\qquad
{\rm with}\qquad \wtd s =\td \lambda e^{-\ft12\epsilon a\phi + dA}\, r^{-(\td
d+1)}\ . 
\ee
This has the solution $e^{-(D-2) \Delta A/(2\td d)} = 1 + \td \lambda 
\sqrt\Delta/(2\td d)\, r^{-\td d}$.  Thus from (\ref{eom1}) we have
\be
2 A' f' = (A'' + \fft{\td d+1}{r} A') (-1 + \fft{\lambda^2}{\td \lambda^2}
e^{-2f})\ ,
\ee
implying
\be
\fft{\lambda^2}{\td \lambda^2} - e^{2f} = c (1 + \fft{\td 
\lambda\sqrt\Delta}{2\td d} r^{-\td d})\ ,
\ee
where $c$ is an integration constant.  Substituting (\ref{fsol}) into this, 
we deduce that
\be
e^{-(D-2) \Delta A/(2\td d)} = 1 + \fft{k}{r^{\td d}}
(\fft{\lambda^2}{\td \lambda^2} -1)^{-1}\ ,
\ee
Thus it is natural to set $\td \lambda= \lambda \tanh\mu$, giving
\be
e^{-(D-2) \Delta A/(2\td d)} = 1 + \fft{k}{r^{\td d}}
\sinh^2\mu .
\ee
The blackened single-scalar $p$-brane solution is therefore given by
\bea
ds^2&=& \Big( 1+ \fft{k}{r^{\td d}} \sinh^2\mu\Big)^{-\fft{4 
\td d}{\Delta(D-2)}} (-e^{2f}dt^2 + dx^idx^i) \nonumber\\
&&+\Big( 1+ \fft{k}{r^{\td d}} \sinh^2\mu\Big)^{\fft{4
d}{\Delta(D-2)}} (e^{-2f} dr^2 + r^2 d\Omega^2)\ ,\nonumber\\
e^{\fft{\epsilon\Delta}{2a} \phi} &=& 1 + \fft{k}{r^{\td d}} \sinh^2\mu
\ ,\qquad e^{2f} = 1 - \fft{k}{r^{\td d}}\ ,\label{sssol}
\eea
with the two free parameters $k$ and $\mu$ related to the charge $\lambda$ 
and the mass per unit $p$-volume, $m$.  Specifically, we find that
\be
\lambda= \fft{\td d k}{\sqrt\Delta} \sinh2\mu\ ,\qquad
m= k \Big(\fft{4\td d}{\Delta} \sinh^2\mu + \td d+1\Big)
\ .\label{mc1}
\ee
The extremal limit occurs when $k \longrightarrow 0$, $\mu \longrightarrow
\infty$ while holding  $k e^{2\mu} =\sqrt\Delta \lambda/\td d={\rm
constant}$. If $k$ is non-negative, the mass and charge satisfy the bound 
\be 
m- \fft{2\lambda}{\sqrt\Delta} =
\fft{k}{\Delta} \Big[(\td d +1) \Delta -2\td d + 
2\td d e^{-2\mu} \Big] \ge \fft{2k \td d^2 (d-1)}{\Delta (D-2)}
\ge 0\ ,\label{bound} 
\ee
where the inequality is derived from $\Delta = a^2 + 2d \td d/(D-2)\ge 2d 
\td d /(D-2)$.  The mass/charge bound (\ref{bound}) is saturated when $k$
goes to zero, which is the extremal limit.  In cases where $\Delta = 4/N$,
the extremal solution becomes supersymmetric, and the bound (\ref{bound})
coincides with the Bogomol'nyi bound.  Note however that in general there
can exist extremal classical $p$-brane solutions for other values of
$\Delta$, which preserve no supersymmetry \cite{lpsol}. 

       It follows from (\ref{temp}) and (\ref{entropy}) that the Hawking
temperature and entropy of the black $p$-brane (\ref{sssol}) are
given by 
\be
T = \fft{\td d}{4\pi r_+} \Big(\cosh\mu\Big)^{-\ft4{\Delta}}\ ,\qquad
S= \ft14 r_+^{\td d+1}\, 
\omega_{\td d+1}\, \Big(\cosh\mu\Big)^{\ft{4}{\Delta}} \ .
\ee
In the extremal limit, they take the form
\be
T \propto (e^\mu)^{2(a^2 - \ft{2\td d^2}{D-2})/(\Delta \td d)}
\ ,\qquad S \propto e^{\mu(4/\Delta-2(\td d+1)/\td d)}
\ .
\ee
Thus the entropy becomes zero in the extremal limit $\mu\rightarrow \infty$,
unless the constant $a$ is zero and $d=1$, since the exponent can be 
rewritten as $\mu(4/\Delta-2(\td d+1)/\td d)=-2\mu\Big(2(d-1)\td d/(D-2) 
+(\td d+1) a^2/\td d)/\Delta$.  In these special cases the dilaton $\phi$
vanishes and the entropy is finite and non-zero.  The situation can arise 
for black holes with $\Delta=4/3$ in $D=5$, and $\Delta=1$ in $D=4$. The
temperature of the extremal $p$-brane is zero, finite and non-zero, or
infinite, according to whether $(a^2 -\fft{2 \td d^2}{D-2})$ is negative,
zero or positive. 

\section{Black dyonic $p$-branes}

     Dyonic $p$-brane occur in dimensions $D=2n$, where the $n$-index field
strengths can carry both electric and magnetic charges.  There are two types
of dyonic solution. In the first type, each individual field strength in
(\ref{mslag}) carries either electric charge or magnetic charge, but not
both.  The construction of black dyonic $p$-branes of this type is identical
to that for the solutions with purely electric or purely magnetic charges,
which we discussed in the previous section. 

     In this section, we shall construct black dyonic $p$-branes of the
second type, where there is one field strength, which carries both electric
and magnetic charge.  The Lagrangian is again given by (\ref{sslag}), with
the field strength now taking the form 
\be
F=  \lambda_1\, \epsilon_n + \lambda_2\, {}*\!\epsilon_{n} \ .\label{dyonf}
\ee
As in the case of purely elementary or purely solitonic $p$-brane solutions,
we impose the conditions (\ref{drama}) and (\ref{feq}) on $B$ and $f$ 
respectively.  The equations of motion then reduce to
\bea
\phi'' + \fft{n}{r} \phi' + 2 \phi' f' &=& \ft12 a (s_1^2 - s_2^2) e^{-2f}
\ ,\nonumber\\
A'' + \fft{n}{r} A' + 2  A' f' &=& \ft14 (s_1^2 + s_2^2) e^{-2f}
\ ,\label{eom2}\\
d(D-2)A'^2 + \ft12 \td d \phi'^2 + 2 (D-2)A' f' &=&
\ft12 \td d (s_1^2 + s_2^2) e^{-2f}\ ,\nonumber
\eea
where
\be
s_1 = \lambda_1 e^{\ft12 a\phi + (n-1) A}\,  r^{-n}\ ,\qquad
s_2 = \lambda_2 e^{-\ft12 a\phi + (n-1) A}\,  r^{-n}\ .
\ee
We can solve the equations (\ref{eom2}) for black dyonic $p$-branes by
following analogous steps to those described in the previous section,
relating the solutions to extremal dyonic solutions.  In particular, we
again find that the functions $A$, $B$ and $\phi$ take precisely the same
forms as in the extremal case, but with rescaled values of charges.
Solutions for extremal dyonic $p$-branes are known for two values of $a$,
namely $a^2 = n-1$ and  $a=0$ \cite{lpsol}.  When $a^2=n-1$, we find that
the black dyonic $p$-brane solution is given by 
\be
e^{-\ft12 a\phi -(n-1) A} = 1 + \fft{k}{r^{n-1}} \sinh^2\mu_1\ ,\qquad
e^{\ft12 a\phi -(n-1) A} = 1 + \fft{k}{r^{n-1}} \sinh^2\mu_2\ ,
\label{dyon}
\ee
with $f$ given by (\ref{fsol}).   The mass per unit volume and the electric
and magnetic charges are given by 
\be
m = k (2\sinh^2\mu_1 + 2\sinh^2\mu_2 + 1)\ ,\qquad
\lambda_\a = (a k/\sqrt2)\,\sinh(2\mu_\a)\ .\label{mc2}
\ee
For the non-negative values of $k$, the mass and the charges satisfy the 
bound
\be
m-(\lambda_1 + \lambda_2)  = k (n-2 + e^{-2\mu_1} + e^{-2\mu_2}) \ge 0\ .
\label{bound2}
\ee
The bound is saturated in the extremal limit $k\longrightarrow 0$.  The
solution (\ref{dyon}) corresponds to the black dyonic string with $n=3$ and
$\Delta=4$ in $D=6$, and the dyonic black hole with $n=2$ and $\Delta=2$ in
$D=4$.  In both cases, the extremal solution is supersymmetric and the bound
(\ref{bound2}) coincides with the Bogomol'nyi bound.  Using (\ref{temp}) and
(\ref{entropy}), we find that the Hawking temperature and entropy of the 
non-extremal solutions are given
by 
\be
T = \fft{\td d}{4 \pi r_+} \Big(\cosh\mu_1 \, \cosh\mu_2\Big)^{-\ft2{n-1}}\ ,
\qquad
S = \ft14 r_+^n\, \omega_n\, \Big(\cosh\mu_1 \cosh\mu_2\Big)^\ft{2}{n-1}\, \ . 
\ee

     When $a=0$, the equations of motion degenerate and the dilaton $\phi$
decouples.  We find the solution 
\be
\phi = 0\ ,\qquad e^{-(n-1)A} = 1 + \fft{k}{r^{n-1}} \sinh^2\mu\ ,
\ee
where again $f$ is given by (\ref{fsol}).  The constant $\mu$ is related to
the electric and magnetic charges by $\sqrt{\lambda_1^2 + \lambda_2^2} = k
\sinh2\mu$.   In this case, unlike the $a^2=n-1$ case, the solution is
invariant under rotations of the electric and magnetic charges, and hence it
is equivalent to the purely electric or purely magnetic solutions we
discussed in the previous section.  Note that in the dyonic solution
(\ref{dyon}), when the parameter $\mu_1 = \mu_2$, {\it i.e.}\ the electric
and magnetic charges are equal, the dilaton field also decouples.  For
example, this can happen if one imposes a self-dual condition on the 3-form
field strength in the dyonic string in $D=6$.  However, this is a different
situation from the $a=0$ dyonic solution, since in the latter case the
electric and magnetic charges are independent free parameters.  In fact the
$a=0$ dyonic solution with independent electric and magnetic charges occurs
only in $D=4$. 

\section{Black multi-scalar $p$-branes}

     To describe multi-scalar $p$-brane solutions, we return to the
Lagrangian (\ref{mslag}) involving $N$ scalars and $N$ field strengths. As
we discussed previously, it can be consistently truncated to the
single-scalar Lagrangian (\ref{sslag}), in which case all the field
strengths $F_\a$ are proportional to the canonically-normalised field
strength $F$, and hence there is only one independent charge parameter.  In
a multi-scalar $p$-brane solution, the charges associated with each field
strength become independent parameters.  After imposing the conditions
(\ref{drama}) and (\ref{feq}), the equations of motion reduce to 
\bea
&&\vp_\a'' + \fft{\td d +1}{r}\vp_\a' + 2 \vp_\a' f' = -\ft12 \epsilon
e^{-2f}\sum_{\beta=1}^N M_{\a\beta}\, S_\beta^2\ , \nonumber\\
&& A'' + \fft{\td d+1}{r} A' + 2 A' f' = \fft{\td d}{2(D-2)} e^{-2f}
\sum_{\a=1}^N S_\a^2\ ,\label{eom3}\\
&&d(D-2)A'^2 + \ft12 \td d \sum_{\a,\beta=1}^N(M^{-1})_{\a\beta}\,
\vp_\a'\, \vp_\beta'+ 2 (D-2)A' f' = \ft12\td d e^{-2f}
\sum_{\a=1}^N S_\a^2\ ,\nonumber
\eea
where $\vp_\a = \vec a_\a \cdot \vec \phi$ and $S_\a = \lambda_\a e^{-\ft12
\epsilon \vp_\a+dA}\, r^{-(\td d +1)}$.  We again find black solutions by
taking $A$ and $\vp_\a$ to have the same forms as in the extremal case, with
rescaled charges.  Extremal solutions can be found in cases where the dot
products of the dilaton vectors $\vec a_\a$ satisfy (\ref{mmatrix})
\cite{lpsol}.  Thus we find that the corresponding black solutions are given
by 
\bea
&&e^{\ft12 \epsilon\vp_\a -dA} = 1 + \fft{k}{r^{\td d}} \sinh^2\mu_\a
\ ,\qquad e^{2f}=1 -\fft{k}{r^{\td d}}\ ,\nonumber\\
&&ds^2 = \prod_{\a=1}^{N} \Big (1 + \fft{k}{r^{\td d}} \sinh^2\mu_\a
\Big)^{-\ft{\td d}{D-2}} ( -e^{2f} dt^2 + dx^i dx^i)\label{mssol}\\
&&\phantom{xxxx} + \prod_{\a=1}^N \Big (1 + \fft{k}{r^{\td d}}
\sinh^2\mu_\a \Big)^{\ft{d}{D-2}} ( e^{-2f} dr^2 + r^2d\Omega^2)\ .\nonumber
\eea
The mass per unit volume and the charges for this solution are given by
\be
m = k (\td d\sum_{\a=1}^N \sinh^2\mu_\a + \td d +1)\ ,
\qquad \lambda_\a= \ft12 \td d k \sinh2\mu_\a\ .
\label{mc3} 
\ee
For non-negative values of $k$, the mass and charges satisfy the bound
\be
m - \sum_{\a=1}^N \lambda_\a = \ft12 k\td d \sum_{\a=1}^N (e^{-2\mu_\a} -1) +
k (\td d + 1) \ge \fft{k \td d (d-1)}{d}\ge 0\ .\label{bound3}
\ee
The bound coincides with the Bogomol'nyi bound. The Hawking temperature and
entropy are given by 
\be
T = \fft{\td d}{4\pi r_+} \prod_{\a=1}^N (\cosh\mu_\a)^{-1}\ ,\qquad S=
\ft14 r_+^{\td d+1}\,  \omega_{\td d+1}\, 
\prod_{\a=1}^{N} \Big(\cosh\mu_\a\Big)\ . 
\ee 
In the extremal limit $k\longrightarrow 0$,  the bound (\ref{bound3}) is
saturated, and the solutions become supersymmetric.

\section{Conclusions}

     We have presented a class of black p-brane solutions of M-theory
which were hitherto known only in the extremal supersymmetric limit and have
calculated their macroscopic entropy and temperature. It would obviously be
of interest to provide a microscopic derivation of the entropy and
temperature using $D$-brane techniques and compare them with the macroscopic
results found in this paper. Agreement would both boost the credibility of
M-theory and further our understanding of black hole and black $p$-brane
physics. 

\section{Acknowledgements}

We have enjoyed useful conversations with Joachim Rahmfeld.  We are very 
grateful to Arkady Tseytlin for drawing our attention to an error in the 
entropy formula (\ref{entropy}) in the original version of this paper.


\begin{thebibliography}{99}

\bibitem{cjs} E. Cremmer, B. Julia and J. Scherk, {\sl Supergravity theory in 
11 dimensions}, Phys. Lett. {\bf B76} (1978) 409.

\bibitem{guven} R. G\"uven, {\sl Black $p$-brane solitons of $D=11$ 
supergravity theory}, Phys. Lett. {\bf B276} (1992) 49.

\bibitem{ds} M.J. Duff and K.S. Stelle, {\sl Multi-membrane solutions of 
$D=11$ supergravity}, Phys. Lett. {\bf B253} 
(1991) 113.

\bibitem{bst} E. Bergshoeff, E. Sezgin and P.K. Townsend, 
{\sl Supermembranes and eleven-dimensional supergravity}, Phys. Lett. {\bf 
B189} (1987) 75; {\sl Properties of the eleven-dimensional supermembrane
theory}, Ann. Phys. {\bf 185} (1988) 330. 

\bibitem{hs} G.T. Horowitz and A. Strominger, {\sl Black strings and 
$p$-branes}, Nucl. Phys. {\bf B360} (1991) 197.

\bibitem{dl} M.J. Duff and J.X. Lu, {\sl Black and super $p$-branes in 
diverse dimensions}, Nucl. Phys. {\bf B416} (1994) 301.

\bibitem{dkrm} M.J. Duff, R. R. Khuri, R. Minasian and J. Rahmfeld,
{\sl New black hole, string and membrane solutions of the four dimensional 
heterotic string},  Nucl. Phys. {\bf B418} (1994) 195.

\bibitem{ht} C.M. Hull and P.K. Townsend, {\sl Unity of superstring 
dualities}, Nucl. Phys. {\bf B294} (1995) 196.

\bibitem{dkl} M.J. Duff, R.R Khuri and J.X. Lu, {\sl String solitons},
Phys. Rep. {\bf 259} (1995) 213.

\bibitem{t2} A. A. Tseytlin, {\sl Extreme dyonic black holes in string theory},
hep-th/9601177.

\bibitem{dr} M.J. Duff and J. Rahmfeld, {\sl Massive string states as 
extreme black holes}, Phys. Lett. {\bf B345} (1995) 441.

\bibitem{sen1} A. Sen, {\sl Black hole solutions in heterotic string theory on a
torus}, Nucl. Phys. {\bf B440} (1995) 421; {\sl Extremal black holes and
elementary string states}, Mod. Phys. Lett. {\bf A10} (1995) 2081; {\sl A note
on marginally stable bound states in Type II string theory}, hep-th/9510229.

\bibitem{km} R. R. Khuri and R. C. Myers, {\sl Dynamics of extreme black
holes and massive string states}, hep-th/9508045.

\bibitem{peet} A. Peet, {\sl Entropy and supersymmetry of D dimensional
extremal electric black holes versus string states}, hep-th/9506200.


\bibitem{dlr} M.J. Duff, J.T. Liu and J. Rahmfeld, {\sl Four dimensional 
string/string/string triality}, Nucl. Phys. {\bf B459} (1996) 125.

\bibitem{r} J. Rahmfeld, {\sl Extremal black holes as bound states},
hep-th/9512089. 

\bibitem{sv} A. Strominger and C. Vafa, {\sl Microscopic origin of 
Bekenstein-Hawking entropy}, hep-th/9601029.

\bibitem{cm} C. Callan and J. Maldacena, {\sl D-brane approach to black hole 
quantum mechanics}, hep-th/9602043.

\bibitem{hs2} G.T. Horowitz and A. Strominger, {\sl Counting states of near 
extremal black holes}, hep-th/9602051.

\bibitem{gm} A. Ghosh and P. Mitra, {\sl Entropy of extremal dyonic black
holes}, hep-th/9602057.

\bibitem{bmpv} J. Beckenridge, R. Myers, A. Peet and C. Vafa, {\sl D-branes 
and spinning black holes}, hep-th/9602065.

\bibitem{dasmath} S.R. Das and S. Mathur, {\sl Excitations of D-strings, 
entropy and duality}, hep-th/9611152.

\bibitem{das} S.R. Das, {\sl Black hole entropy and string theory},
hep-th/9602172.

\bibitem{blmpsv} J. Beckenridge, D. A. Lowe, R. Myers, A. Peet, A. Strominger and
C. Vafa, {\sl Macroscopic and microscopic entropy of near-extremal spinning
black holes}, hep-th/9603078.

\bibitem{sm} A. Strominger and J. Maldacena, {\sl Statistical entropy of
four-dimensional extremal back holes}, hep-th/9603060. 

\bibitem{jkm} C.V. Johnson, R.R. Khuri and R.C. Myers, {\sl Entropy of 4-$D$ 
extremal black holes}, hep-th/9603061.


\bibitem{hms} G. T. Horowitz, J. Maldacena and A. Strominger, {\sl Nonextremal
black hole microstates and U-duality}, hep-th/9603109.

\bibitem{dvv} R. Dijkgraaf, E. Verlinde and H. Verlinde, {\sl BPS spectrum of
the five-brane and black hole entropy}, hep-th/96033126. 

\bibitem{p} J. Polchinski, {\sl Dirichlet-branes and Ramond-Ramond charges}, 
hep-th/9511026.

\bibitem{lpsol} H. L\"u and C.N. Pope, {\sl $p$-brane solitons in maximal 
supergravities}, hep-th/9512012, to appear in Nucl. Phys. {\bf B}.

\bibitem{lpmulti} H. L\"u and C.N. Pope, {\sl Multi-scalar $p$-brane
solitons}, hep-th/9512153, to appear in Int. J. Mod. Phys. {\bf A}.

\bibitem{lpss} H. L\"u, C.N. Pope, E. Sezgin and K.S. Stelle, {\sl Stainless 
super $p$-branes}, Nucl. Phys. {\bf B456} (1995) 669.

\bm{ght} G.W. Gibbons, G.T. Horowitz and P.K. Townsend, {\sl Class. Quantum
Grav.} {\bf 12} (1995) 297.

\bibitem{lw} F. Larsen and F. Wilczek, {\sl Internal structure of black holes},
hep-th/9511064.

\bibitem{dp}  M.J. Duff and C. N. Pope, {\sl Consistent truncations in
Kaluza-Klein theories}, Nucl. Phys. B255 (1985) 355.

\bibitem{po} C.N. Pope, {\sl The embedding of the Einstein Yang-Mills 
equations in $D=11$ supergravity}, Class. Quant. Grav. {\bf 2} (1985) L77.

\bibitem{dl2} M.J. Duff and J.X. Lu, {\sl The self-dual type IIB 
superthreebrane}, Phys. Lett. {\bf B273} (1991) 409.

\bibitem{cy} M. Cvetic and D. Youm, {\sl Dyonic BPS saturated black holes of
heterotic string theory on a torus}, hep-th/9507090.

\bibitem{pt} G. Papadopoulos and P. K. Townsend, {\sl Intersecting $M$-branes},
hep-th/9603087.

\bibitem{t1} A. A. Tseytlin, {\sl Harmonic superpositions of $M$-branes},
hep-th/9604035.

\bibitem{dfkr} M.J. Duff, S. Ferrara, R.R. Khuri and J.
Rahmfeld, {\sl Supersymmetry and dual string solitons}, Phys. Lett. {\bf B356}
(1995) 479.

\bibitem{lpx} H. L\"u, C.N. Pope and K.W. Xu, {\sl Liouville and Toda 
solitons in M-theory}, hep-th/9604058.

\bibitem{l} J.X. Lu, {\sl ADM masses for black strings and $p$-branes},
Phys. Lett. {\bf B313} (1993) 29.



\end{thebibliography}
\end{document}